\documentclass[prb,twocolumn,amsmath,amssymb,letterpaper,superscriptaddress,nofootinbib]{revtex4-2} 
\usepackage{graphicx}
\usepackage{amsmath}
\usepackage{amsfonts}
\usepackage{amssymb}
\usepackage{braket}
\usepackage[colorlinks,linkcolor=blue,filecolor=blue,urlcolor=blue,citecolor=blue]{hyperref}

\begin{document}

\title{Biorthogonal topological charge pumping in non-Hermitian systems}

\author{Zhenming Zhang}
\affiliation{CAS Key Laboratory of Quantum Information, University of Science and Technology of China, Hefei 230026, China}
\author{Tianyu Li}
\affiliation{Beijing National Laboratory for Condensed Matter Physics, Institute of Physics, Chinese Academy of Sciences, Beijing 100190, China}
\author{Xi-Wang Luo}
\email{luoxw@ustc.edu.cn}
\affiliation{CAS Key Laboratory of Quantum Information, University of Science and Technology of China, Hefei 230026, China}
\affiliation{CAS Center For Excellence in Quantum Information and Quantum Physics, Hefei 230026, China}
\author{Wei Yi}
\email{wyiz@ustc.edu.cn}
\affiliation{CAS Key Laboratory of Quantum Information, University of Science and Technology of China, Hefei 230026, China}
\affiliation{CAS Center For Excellence in Quantum Information and Quantum Physics, Hefei 230026, China}

\begin{abstract}
We study charge pumping in generic non-Hermitian settings, and show that quantized charge pumping is only guaranteed under a biorthogonal formalism therein, where the charge transport is evaluated using the left and right eigenvectors of the non-Hermitian Hamiltonian.
Specifically, for biorthogonal charge pumping in generic one-dimensional non-Hermitian models, we demonstrate how
quantized transport is related to the Chern number in the parameter space. When the non-Hermitian model possesses the non-Hermitian skin effect, under which
Bloch states in the bulk are deformed and localize toward boundaries,
we propose a scenario where the pumped charge is related to
the non-Bloch Chern number defined in the parameter space involving the generalized Brillouin zone.
We illustrate the validity of our analytic results using concrete examples, and, in the context of the biorthogonal charge pumping, discuss in detail a recent experiment where quantized charge pumping was observed in a lossy environment.
\end{abstract}

\maketitle

\section{Introduction}

Topological charge pumping, first considered by Thouless in 1983, describes the quantized transport of particles along a one-dimensional lattice potential under slow but cyclic drive~\cite{Thouless1983, NiuThouless}. Since the transported charge per driving cycle is dictated by a Chern number defined on a $(1+1)$D momentum-time manifold, the phenomenon is intrinsically topological, and serves as an important paradigm in the study of topological matter~\cite{tknn,toporev1,toporev2}.
To date, Thouless pumping has been experimentally observed in synthetic configurations such as photonics and waveguide arrays~\cite{waveguide1, waveguide2, waveguide3, waveguide4, waveguide5}, and ultracold atoms in optical lattice potentials~\cite{ultracold1, ultracold2, ultracold3}.
Its many variants, such as the quantized transport under fast cyclic drive~\cite{fast1, fast2,fast3,fast4} (as in the context of Floquet topological matter), or that under dissipation in open systems~\cite{nc, dissipative}, have also attracted much attention.
Particularly, since physical systems are inevitably coupled to their environment, whether quantized charge pumping persists in the presence of dissipation is an important issue, and closely related to potential applications in open systems.

In this work, we study charge pumping in generic one-dimensional non-Hermitian models, in an effort to unveil its quantization condition.
The non-Hermiticity discussed here signifies the intrinsic openness of the system under study~\cite{molmer,weimer}, which can be enforced either by introducing gain and loss in classical settings~\cite{photonics3,photonics4}, through post selection in the dynamics of open quantum systems~\cite{Non1,uedareview}, or, more generally, by vectorizing the open-system density matrix so that its Liouvillian dynamics is mapped to that driven by a non-Hermitian effective Hamiltonian~\cite{mastereqeff1,mastereqeff2,zhushiliang,tianyu}. Without loss of generality, here we focus on the dynamics under a non-Hermitian Hamiltonian, regardless of its physical origin.

In the standard Thouless pumping, the quantized particle transport arises for a fully-filled band under a periodic modulation of the corresponding one-dimensional lattice potential in the adiabatic regime. The adiabaticity requires that the time scale of the modulation be much smaller than that associated with the band gap.
On extending the scenario to the non-Hermitian case, two crucial questions naturally arise. First, given that the right eigenstates of non-Hermitian Hamiltonians are non-orthonormal in general, the charge transport per cycle may no longer be quantized. This is because the charge transfer is related to the expectation value of the position operator, where the orthonormality between eigenstates in the same band is crucial~\cite{Non1}.
Second, eigenstates of non-Hermitian systems often acquire complex eigenvalues with distinct imaginary components. The exponentially different growth of the eigenstates during the non-unitary time evolution can be detrimental to the adiabatic condition~\cite{nhadia}.
Furthermore, in non-Hermitian topological systems, the introduction of open boundaries can give rise to the non-Hermitian skin effect~\cite{nhtopot4,nhtopot5,murakami,nhse1,nhse2,nhse3,nhse4,nhse5,nhse6,nhsedy1,nhsedy2,nhsedy3,skinrev1,skinrev2}, deforming the Bloch states in the bulk and necessitating the so-called non-Bloch topological invariants to account for the band topology under the open boundary condition~\cite{nhtopot4,nhtopot5,murakami}. It is then tempting to
ask whether a dynamic pumping process exists where the quantized transport corresponds to a non-Bloch topological invariant following the non-Bloch band theory.

We propose to address the questions above through the formulation of biorthogonal transport, where a quantized charge pumping is ensured through the biorthonormality of the left and right eigenstates of the non-Hermitian model. This is inspired by existing studies where biorthogonal formalisms are adopted to study
the anomalous boundary modes and localization in non-Hermitian models~\cite{emilskin,TNag,emilapl,mandal}, the quantized charge pumping in pseudo-Hermitian systems under the periodic boundary condition~\cite{entropy}, and the non-Bloch topological invariants~\cite{xiwang,nhsedy2,quenchexp}.
Here, regardless of the boundary condition, we consider a general charge pumping procedure where the system is initialized in a fully-filled band, and subject to periodic modulation in the parameter space.
We discuss the adiabatic condition in non-Hermitian pumping processes, arguing that an approximate adiabaticity can be achieved at intermediate timescales.
Under these conditions, we demonstrate that a
quantized charge pumping is guaranteed under the biorthogonal formulation, which is related to a Chern number defined in the parameter space. In the presence of the non-Hermitian skin effect and under the open boundary condition, this gives rise to a direct correspondence between the quantized charge transfer and the non-Bloch Chern number.
The formalism of biorthogonal charge pumping thus complements previous discussions of biorthogonal chiral displacement and non-Bloch quench~\cite{xiwang,nhsedy2,quenchexp}, and further highlights the utility of non-Bloch band theory in describing the dynamic topological phenomena in the non-Hermitian setting.
We then confirm the validity of our general analysis using concrete examples. Importantly, quantized non-Hermitian charge pumping was recently observed in plasmonic waveguide arrays without invoking the biorthogonal formalism. We show that the experimentally observed quantized charge pumping is only approximate. It derives from particular features of the implemented Hamiltonian, and is therefore not applicable to more general non-Hermitian models. Based on existing experimental probe of the biorthogonal chiral displacement~\cite{quenchexp}, we expect that the biorthogonal charge pumping can be demonstrated in both quantum and classical systems~\cite{classical exp 1,classical exp 2,classical exp 3}, where non-Hermitian dynamics have been engineered and controlled.

The work is organized as follows. In Sec.~II, we review the standard Thouless pumping in the Hermitian case. We then present the general framework of biorthogonal pumping in Sec.~III. In Sec.~IV, we discuss concrete examples, particularly in the light of a recent experiment on non-Hermitian pumping. We summarize in Sec.~V.

\section{Thouless pumping in the Hermitian case}\label{sec:H}

In the standard Thouless pumping, a fully-filled band of a one-dimensional dimerized chain (the Rice-Mele model~\cite{RM}) is subject to a cyclic but slow modulation of the Hamiltonian. Without loss of generality, in the following discussion, we consider a tight-binding Hamiltonian $\hat{H}(t)$ with time-dependent parameters. The instantaneous bulk eigenstates are given as
\begin{align}
 \ket{\Psi_{k}(t)}=\frac{1}{\sqrt{N}}\sum_m e^{imk}\ket{m}\otimes \ket{u_{k}(t)},
\end{align}
where $k\in [0, 2\pi)$ is the quasimomentum in the first Brillouin zone (BZ), $m$ is the unit-cell index, $N$ is the total number of unit cells, and $\ket{u_{k}(t)}$ is the eigenstate of the Bloch Hamiltonian $\hat{H}(k,t)$, with $\hat{H}(k,t)\ket{u_k(t)}=\epsilon_k(t)\ket{{u}_{k}(t)}$ and $\epsilon_k(t)$ is the dispersion. Here the full Hilbert space is spanned by $|m,s\rangle=|m\rangle\otimes |s\rangle$, where $s$ is the sub-lattice index (two dimensional for the dimerized chain). It follows that $|u_k\rangle$ represents the wavefunction within a unit cell, and $|k\rangle=\frac{1}{\sqrt{N}}\sum_m e^{ikm}|m\rangle$ gives the plane-wave state in the space spanned by $\{|m\rangle\}$. In the following, we refer to $m$ as the site index, and $s$ as the index of the internal degrees of freedom on each site.
We omit the band index for the fully-filled band, but for any other band, we have $\hat{H}(k,t)\ket{u_{k,n}(t)}=\epsilon_{k,n}(t)\ket{u_{k,n}(t)}$,
where $n$ is the band index, and $\epsilon_{k,n}$ and $\ket{u_{k,n}(t)}$ are the corresponding band dispersion and eigenstate, respectively.

The particle transfer (or the pumped charge) over one cycle of parameter modulation (of the period $T$) is given by the average displacement of the time-evolved band, which can be written as
\begin{equation}
    \Delta \bar x = \int_0^T  dt ~\partial_t \bar{x}(t),
\end{equation}
where $\bar x(t)$ is the average position of the band, defined as
\begin{equation}
    \bar{x}(t) :=  \int_{\text{BZ} } \frac{dk}{2\pi} \langle \tilde\Psi_k(t)|\hat{x}|\tilde\Psi_k(t)\rangle.\label{eq:barxdef}
\end{equation}
Here $|\tilde\Psi_k(t)\rangle $ represents the time-evolved state, which is initialized as the eigenstate of the Hamiltonian at $t=0$, and driven by $\hat{H}(t)$. It can be written in the Bloch-like form as
\begin{equation}
    |\tilde\Psi_k(t)\rangle = \frac{1}{\sqrt{N}}\sum_m e^{imk}|m\rangle\otimes |\tilde u_k(t)\rangle,
    \end{equation}
where $|\tilde u_k(t)\rangle$ is initialized as $| u_k(0)\rangle$ and driven by the Bloch Hamiltonian $\hat{H}(k,t)$. According to the Schr\"{o}dinger equation, we obtain
\begin{equation}
    \partial_t \bar{x}(t) = -i \int_{\text{BZ}} \frac{dk}{2\pi} \langle \tilde\Psi_k(t)|[\hat{x},\hat{H}(t)]|\tilde\Psi_k(t)\rangle,\label{eq:SE}
\end{equation}
and equivalently,
\begin{equation}
     \partial_t \bar{x}(t) = \int_{\text{BZ}} \frac{dk}{2\pi} \bra{\tilde{u}_k(t)} \frac{\partial \hat{H}(k,t) }{\partial k}\ket{\tilde{u}_k(t)}.\label{eq:Dx1}
\end{equation}

Under the time-dependent perturbation, the time-evolved internal state is given by
\begin{align}
   \ket{\tilde{u}_{k}(t)}\approx e^{i\kappa(k,t)}  \ket{{u}_{k}(t)}+ \sum_n a_n \ket{{u}_{k,n}(t)} ,\label{eq:adcond}
\end{align}
where the phase factor $\kappa(k,t)$ contains both the geometric- and dynamic-phase contributions
$\kappa(k,t) = \int_0^t d\tau [i\left\langle{u_k(\tau)}|{\partial_\tau u_k(\tau)}\right\rangle-\epsilon_k(\tau)]$.
We choose the parallel-transport gauge in which $ \langle u_k(\tau)|\partial_\tau u_k(\tau)\rangle = 0$, so that $\kappa(k,t)$ only contains the dynamic phase.
It follows that $a_n$ is given by
\begin{equation}\label{eq:expan}
    a_n = i e^{i\kappa(k,t)} \frac{\langle u_{k,n}(t)|\partial_t |u_k(t)\rangle }{\epsilon_{k,n}-\epsilon_k}.
\end{equation}

In Eq.~(\ref{eq:Dx1}), we identify the total velocity in each $k$ sector
$v_k:=\bra{\tilde{u}_k(t)} [\partial_k \hat{H}(k,t) ]\ket{\tilde{u}_k(t)}$, which is given by
\begin{align}
v_k&=\frac{\partial \epsilon_k(t)}{\partial k}-i(\langle\partial_k {u}_k(t)|\partial_t {u}_k(t)\rangle-\langle \partial_t {u}_k(t)|\partial_k {u}_k(t)\rangle)\nonumber \\
&:=\frac{\partial \epsilon_k(t)}{\partial k}+\Omega_{kt}. \label{eq:omega}
\end{align}
Here the first term is the group velocity, and the second is the anomalous contribution from the Berry curvature $\Omega_{kt}$ of the Hamiltonian $\hat{H}(k,t)$, defined over the (1+1)D momentum-time manifold.
Since the integral of the group velocity over the momentum-time manifold vanishes, it follows that the total charge pumped over one cycle is equal to the quantized Chern number $C$
\begin{equation}\label{Hermitiangvpump}
   \Delta \bar x=\int_{0}^{T} d t \int_{\text{BZ}} \frac{dk}{2\pi}~v_k =\int_{0}^{T} d t \int_{\text{BZ}} \frac{d k}{2 \pi}\Omega_{kt} =C.
\end{equation}

An alternative description of the Thouless pumping is through the displacement of the Wannier center over one cycle. Here the Wannier states are defined as follows
\begin{equation}
    \ket{w_j(t)} := \frac{1}{\sqrt N}\sum_{k\in \text{BZ}}e^{-ijk}\ket{\tilde \Psi_k(t)}.
\end{equation}
According to Eq.~(\ref{eq:barxdef}), we have
\begin{equation}
    \bar{x}(t) = \frac{1}{N}\sum_j \langle w_j(t) |\hat{x} |w_j(t)\rangle.
\end{equation}
Here we further identify the time-evolved Wannier center of site $j$ as $x_j(t) :=  \left\langle w_{j}(t)|\hat{x}| w_{j}(t)\right\rangle$, which, to the leading order of the time-dependent perturbation, is
\begin{equation}\label{HermitianWannierpump}
  x_j(t)=j+\frac{i}{2 \pi} \int d k\left\langle {u}_{k }(t) | \partial_{k} {u}_{k }(t)\right\rangle.
\end{equation}
Noting that the Wannier centers of different sites only differ by a time-independent site index $j$, we have
\begin{equation}
\Delta x_j=   \int_0^T dt~\partial_t x_j(t) =\int_0^T dt\int_{\text{BZ}}\frac{dk}{2\pi}\Omega_{kt}=C, \label{eq:Hdwc}
\end{equation}
recovering the result of $\Delta \bar x$ in Eq.~(\ref{Hermitiangvpump}).

Note that the above discussions naturally assume the periodic boundary condition (PBC).
In the Hermitian case, charge pumping in the bulk under the open boundary condition (OBC) is rarely discussed explicitly, since, aside from a small number of edge modes, the bulk eigenstates and eigenenergies are insensitive to the boundary conditions in the thermodynamic limit. Hence, regardless of the boundary conditions, one may formally resort to a unified prescription of the Thouless pumping, that is, the average displacement of all eigenstates in a fully-filled band
\begin{align}
\Delta \bar{x}=\bar{x}(T)-\bar{x}(0),
\end{align}
where $\bar{x}(t)=\sum_{i}\bra{\Psi_i(t)}\hat{x}\ket{\Psi_i(t)}/N$ and $i$ is the state label.
Under the OBC, the summation runs over all time-evolved eigenstates $|\Psi_i\rangle$ in the said band, and $\Delta\bar{x}$ approaches the Chern number in the thermodynamic limit.
Under the PBC, the summation runs over all quasimomenta in the BZ, and $\Delta\bar{x}$ recovers the Chern number exactly.
As we show below, such a unified prescription forms the basis for quantized charge transport in non-Hermitian settings where the boundary condition may affect the bulk states.

\section{Biorthogonal pumping in non-Hermitian settings}\label{sec:NHpump}
In this section, we extend the Thouless pumping discussed above to non-Hermitian systems. Two remarks are in order before we start.

First, the Thouless pumping theory is based on adiabatic driving. This is straightforward in Hermitian systems with a finite band gap. However, the situation becomes complicated in non-Hermitian systems, where the eigenspectrum, along with the band gap, becomes complex.
Here the adiabatic approximation corresponds to the assumption that an eigenstate in a given band only acquires a phase factor during the time evolution.
Similar to the Hermitian case, a necessary condition is that the system parameter cannot vary too fast compared to the real component of the band gap.
However, the existence of the imaginary component of the band gap may still lead to deviations from adiabaticity in the long-time limit.

Specifically, for a general non-Hermitian time-dependent Hamiltonian $\hat{H}(t)$, there exist instantaneous left and right eigenstates $\ket{\psi_n^{L,R}(t)}$, which satisfy $\hat{H}(t)\ket{\psi_n^R(t)}=E_n\ket{\psi_n^R(t)}$ and $\hat{H}^\dag\ket{\psi_n^L(t)}=E_n^*\ket{\psi_n^L(t)}$.
In the spirit of Eq.~(\ref{eq:expan}), the adiabatic condition for the initial right eigenstate $\ket{\psi_n^R(0)}$ (under the drive of $\hat H(t)$) is given by~\cite{nhadia}
\begin{equation}
    \frac{|\langle {\psi}_n^L(t) | \dot{\psi}_m^R(t)\rangle|}{\left|\omega_{n m}(t)\right|} \exp[{-\operatorname{Im}\int_0^t  \omega_{n m}(\tau)d\tau  }] \ll 1\label{eq:condition}
\end{equation}
for $m\neq n$, where $\omega_{nm}=E_n-E_m$. Note that in non-Hermitian models with complex dispersions, it is the exponential factor in Eq.~(\ref{eq:condition}) that determines the adiabaticity under a sufficiently long evolution time. More specifically, for a non-Hermitian pumping process, the adiabatic condition does not necessarily imply an infinitely long evolution time. From a practical point of view, if the driving period $T$ is appropriate, such that
Eq.~(\ref{eq:condition}) is satisfied, the dynamics can then be regarded as in the adiabatic regime.
Furthermore, Eq.~(\ref{eq:condition}) is sufficient to ensure adiabaticity of the right eigenstates and hence useful for calculating time-dependent expectation values involving only time-evolved right eigenstates. However, for the biorthogonal formalism that we introduce in this section,
expectation values are taken using both the time-evolved left and right eigenstates.
Since the eigenvalues of $\hat{H}^\dagger(t)$ are complex-conjugate to those of $\hat{H}(t)$, in the long-time limit, the adiabatic condition Eq.~(\ref{eq:condition}) does not simultaneously apply to the evolution of the left and right eigenstates, provided the corresponding eigenvalues are complex.
In this case, the only viable option is the intermediate time, where the adiabatic condition can be approximately satisfied by both the left and right eigenstates.
We will discuss in more detail the manifestations of these conditions with concrete examples in Sec.~\ref{sec:example}.
In this section, we only consider systems and pumping processes satisfying the approximate adiabatic condition.

The second remark is related to boundary conditions.
A noteworthy feature of non-Hermitian systems is the potential impact of boundary conditions on the bulk eigenstates through the non-Hermitian skin effect. Hence, it is expected that different boundary conditions may call for distinct formulation of charge pumping in non-Hermitian systems.

\subsection{Pumping under the PBC}\label{nHPBC}
We first discuss pumping under the PBC, where the quasimomenta $k$ are still good quantum numbers.
While the general charge-pumping formalism connects the total velocity and average displacement respectively with the Berry curvature and Chern number, there is however one important difference. A non-Hermitian Hamiltonian $\hat{H}(t)$ features distinct left and right eigenstates,
which are biorthonormal.
As a result, the instantaneous left and right bulk eigenstates are
\begin{equation}
    \left|\Psi_{k}^{L, R}(t)\right\rangle =\frac{1}{\sqrt{N}} \sum_{m} e^{imk} \ket{m}\otimes \ket{u_k^{L,R}(t)},\label{eq:eigenPBC}
\end{equation}
where the left and right eigenstates of the Bloch Hamiltonian respectively satisfy $\hat{H}(k,t)|u^R_k(t)\rangle=\epsilon_k(t)|u^R_k(t)\rangle$ and $\hat{H}^\dag(k,t)|u^L_k(t)\rangle=\epsilon^\ast_k(t)|u^L_k(t)\rangle$, now with complex $\epsilon_k$.
It follows that, one may define the biorthogonal Berry curvatures
\begin{align}
\Omega^{\alpha\beta}_{kt}=-i(\langle\partial_k {u}_k^{\alpha}(t)|\partial_t {u}^\beta_k(t)\rangle-\langle \partial_t {u}^\alpha_k(t)|\partial_k {u}^\beta_k(t)\rangle),\label{eq:defBerrycurv_PBC}
\end{align}
with the normalization condition $\langle u_k^{\alpha}|u_k^{\beta}\rangle =1$, where $\alpha,\beta\in \{L,R\}$. Previous studies have pointed out that, while these Berry curvatures can be different in general, when integrated over the momentum-time manifold, they would give the same Chern number~\cite{nhchernPBC}. For instance, we have
\begin{align}
C=\int_{0}^{T} d t \int_{\text{BZ}} \frac{d k}{2 \pi}\Omega^{RR}_{kt}=\int_{0}^{T} d t \int_{\text{BZ}} \frac{d k}{2 \pi}\Omega^{LR}_{kt}.
\end{align}

However, the presence of biorthogonal eigenstates also gives rise to different ways of defining the average position
\begin{equation}
    \bar{x}^{\alpha\beta}(t) :=  \int_{\text{BZ} } \frac{dk}{2\pi} \langle \tilde\Psi^{\alpha}_k(t)|\hat{x}|\tilde\Psi^{\beta}_k(t)\rangle,
\end{equation}
where the time-evolved left and right states $|\tilde{\Psi}_k^{R}(t)\rangle$ and $|\tilde{\Psi}_k^{L}(t)\rangle$ are driven by $\hat{H}(t)$ and $\hat{H}^\dagger(t)$, respectively.
Without loss of generality, here we mainly discuss the biorthogonal form $\bar x^{LR}$ and the conventional one $\bar x^{RR}$.

Following the discussion in Sec.~\ref{sec:H},
we first examine the total velocity. Using the Schr\"{o}dinger equation, we obtain
\begin{equation}
    \partial_t \bar{x}^{LR}(t) = \int_{\text{BZ}} \frac{dk}{2\pi} \langle \tilde u_k^L(t) | [\partial_k \hat{H}(k,t)] | \tilde u_k^R(t)\rangle,
\end{equation}
where the time-evolved left and right states $|\tilde{u}_k^{R}(t)\rangle$ and $|\tilde{u}_k^{L}(t)\rangle$ are driven by $\hat{H}(k,t)$ and $\hat{H}^\dagger(k,t)$, respectively, and satisfy $\ket{\tilde u_k^{L,R}(0)} = \ket{ u_k^{L,R}(0)} $.

Similarly to the Hermitian case, we define the biorthogonal total velocity as
\begin{equation}
 v^{LR}_k(t) := \langle \tilde u_k^L(t) | [\partial_k \hat{H}(k,t)] | \tilde u_k^R(t)\rangle.\label{eq:gv}
\end{equation}
Assuming the evolution is approximately adiabatic, we have (using the same convention on the band index $n$)
\begin{align}
|\tilde{u}_k^{R}(t)\rangle&\approx e^{i\kappa(k,t)}|{u}_k^{R}(t)\rangle+\sum_n a^R_n |{u}_{k,n}^{R}(t)\rangle,\label{eq:1orderR}\\
|\tilde{u}_k^{L}(t)\rangle&\approx e^{i\kappa^*(k,t)}|{u}_k^{L}(t)\rangle+\sum_n a^L_n |{u}_{k,n}^{L}(t)\rangle\label{eq:1orderL},
\end{align}
where
$\kappa(k,t)  = \int_0^t d\tau [i\left\langle{u_k^L(\tau)}|{\partial_{\tau}{u}_k^R(\tau)}\right\rangle-\epsilon_k(\tau)]$,
and $\hat{H}(k,t)|u^R_{k,n}(t)\rangle = \epsilon_{k,n}|u^R_{k,n}(t)\rangle$, $\hat{H}^\dagger (k,t)|u^L_{k,n}(t)\rangle = \epsilon^*_{k,n}|u^L_{k,n}(t)\rangle$. We choose the parallel-transport gauge in which $\langle u_k^L(t)|\partial_t u_k^R(t)\rangle = 0$, and it follows that $\kappa(k,t) = -\int_o^T d\tau \epsilon_k(\tau)$. The coefficients $a^{L,R}_n$ are given by
\begin{align}
    a^R_n &= ie^{i\kappa(k,t)}\frac{\langle u_{k,n}^L(t)|\partial_t u_k^R(t)\rangle }{\epsilon_{k,n}(t)-\epsilon_{k}(t)},\\
    a^L_n &= ie^{i\kappa^*(k,t)}\frac{\langle u_{k,n}^R(t)|\partial_t u_k^L(t)\rangle }{\epsilon^*_{k,n}(t)-\epsilon^*_{k}(t)}.
\end{align}
Substituting Eqs.~(\ref{eq:1orderR}) and (\ref{eq:1orderL}) into Eq.~(\ref{eq:gv}), we see that the biorthogonal velocity $v^{LR}_k(t)$ relates to the Berry curvature in a fashion similar to Eq.~(\ref{eq:omega}).
Namely, we have
\begin{align}
   v^{LR}_k(t)=\frac{\partial\epsilon_k(t)}{\partial k}+\Omega^{LR}_{kt}.
\end{align}
Since the group velocity $\frac{\partial\epsilon_k(t)}{\partial k}$ vanishes when integrated over
the momentum-time manifold, the biorthogonal displacement $\Delta \bar{x}^{LR}=\int_0^T dt\int_{\text{BZ}}\frac{dk}{2\pi}v_k^{LR}$ is quantized and recovers the Chern number.

Equivalently, we can also construct a quantized pumping through the Wannier center by adopting the biorthogonal formalism.
Specifically, we define the time-evolved biorthogonal Wannier states
\begin{align}
\ket{\tilde{w}_j^{L,R}(t)}=\frac{1}{\sqrt{N}} \sum_{k} e^{-i j k}\left|{\tilde\Psi}_{k}^{L, R}(t)\right\rangle,
\end{align}
where $|\tilde{\Psi}_k^{R}(t)\rangle =e^{i\kappa(k,t)}|{\Psi}_k^{R}(t)\rangle$ and $|\tilde{\Psi}_k^{L}(t)\rangle =e^{i\kappa^*(k,t)}|{\Psi}_k^{L}(t)\rangle$.
According to the definition of the biorthogonal Wannier states, the average displacement can be equivalently expressed as the average Wannier center
\begin{equation}
    \bar{x}^{LR}(t)= \frac{1}{N} \sum_j \langle \tilde w_j^L(t) |\hat{x} |\tilde w_j^R(t)\rangle .
\end{equation}
The time-evolved biorthogonal Wannier center is then
\begin{align}
     x_j^{LR}(t)&:=  \langle{\tilde{w}_j^{L}(t)}|\hat{x}|{\tilde{w}_j^{R}(t)}\rangle\nonumber\label{eq:wc_LR_PBC}\\
  &= j+  i\int_{\text{BZ}} \frac{dk}{2\pi}\bra{u_k^L(t)}{\partial_k u_k^R(t)}\rangle.
 \end{align}
It is straightforward to show that the shift of the Wannier center, evaluated through the biorthogonal Wannier states, is quantized and equal to the Chern number
\begin{align}
 \Delta x^{LR}_j =  \int_0^T dt~ \partial_t x_j^{LR}(t) =C.
\end{align}

However, when discussing the conventional case $\bar x^{RR}$, a prime issue arises: since the band dispersion generally acquires imaginary components in non-Hermitian models, the norm of $\ket{\tilde{\Psi}_k^R(t)}$ (or of $\ket{\tilde{u}_k^R(t)}$) suffers from an exponential growth or decay, and cannot be canceled out as in the case of the biorthogonal formalism.
Specifically, $\partial_t \bar x^{RR}(t)$ corresponds to the unnormalized Berry curvature, and its integral over the momentum-time manifold is not quantized. To avoid this problem, we can introduce the average renormalized position:
\begin{align}
    \bar{x}^{\text{re}}(t)&:=\frac{1}{2\pi}\int_{\text{BZ}}dk  \frac{\bra{\tilde{\Psi}^R_k(t)}\hat{x}\ket{\tilde{\Psi}^R_k(t)}}{\langle\tilde{\Psi}^R_k(t)|\tilde{\Psi}^R_k(t)\rangle}\nonumber\\
    &= \frac{1}{2\pi}\int dk ~{}_{\text{re}}\langle{\tilde{\Psi}_k^R(t)}|\hat{x}|{\tilde{\Psi}_k^R(t)}\rangle_{\text{re}},
\end{align}
where, under adiabatic assumption, $\ket{\tilde{\Psi}_k^R(t)}_{\text{re}}=e^{i\text{Re}\kappa(k,t)}\ket{\Psi_k^R(t)}$. It follows that
\begin{equation}\label{eq:rext}
    \partial_t \bar x^{\text{re}}(t) = \int_{\text{BZ}} \frac{dk}{2\pi} \Omega_{kt}^{RR},
\end{equation}
which means that the quantized pumping can be recovered by introducing the average renormalized position.
However, Eq.~(\ref{eq:rext}) is not equivalent to the total velocity or the Wannier-center shift in the non-Hermitian case.

More concretely, following the derivations leading to Eq.~(\ref{eq:SE}), we have
\begin{equation}
    \partial_t \bar{x}^{RR} = i\int_{\text{BZ}}\frac{dk}{2\pi} \langle \tilde \Psi_k^R(t)|[\hat{H}^\dagger(t) \hat{x} - \hat{x} \hat{H}(t)]| \tilde \Psi_k^R(t)\rangle.\label{eq:vrr_fail_reason}
\end{equation}
Due to the non-Hermiticity of the Hamiltonian, it cannot be equivalently represented by the total velocity calculated through the right eigenstates. Therefore, the connection between the velocity and $\partial_t \bar{x}^{RR}$, as in Eq.~(\ref{eq:Dx1}), is lost.
Similarly, one can show that the average renormalized position defined above cannot be connected to the total velocity (calculated through the right eigenstates) either.

For the time-evolved Wannier center without resorting to
the biorthogonal formalism, we have
\begin{equation}
    x_j^{RR}(t) :=\frac{\langle \tilde{w}_j^R(t)|\hat{x}|\tilde{w}_j^R(t)\rangle}{\langle \tilde{w}_j^R(t)|\tilde{w}_j^R(t)\rangle}. \label{eq:wc_RR_PBC}
\end{equation}
Note that the initial right Wannier state $|\tilde{w}_j^R(0)\rangle$
is an equal weight superposition (with a phase factor) of the right Bloch states. Under the adiabatic time evolution driven by the non-Hermitian Hamiltonian, however, $\text{Im}\epsilon_k$ at different quasimomentum $k$ would cause exponential growth or decay of the relative weight in the superposition. After one cycle, the right Wannier state would be dominated by the Bloch state with the largest imaginary eigenenergy component. This is the reason why $\Delta x_j^{RR}$ fails to recover the quantized Chern number here. Nevertheless, if the imaginary part of the band dispersion is flat (or nearly flat), then $x_j^{RR}(t)$ will be equivalent to (or nearly equivalent to) $\bar{x}^{\text{re}}(t)$, leading to a quantized pumping. An example of this particular case is given in Sec.~\ref{sec:modelB}.

\subsection{Pumping under the OBC}\label{sec:OBC}
We now consider charge pumping under the OBC in a non-Hermitian system, where the boundary effect can play a significant role. For this purpose, we consider the time evolution of a fully-filled band under the OBC, where the occupied left and right time-evolved states are denoted as $\ket{\tilde{\Psi}_i^{L, R}(t)}$, with $i$ being the state index.
Similar to the previous cases, $\ket{\tilde{\Psi}_i^{L, R}(t)}$ are respectively the left and right eigenstates of the Hamiltonian at $t=0$ (with the normalization condition $\langle \Psi_i^L | \Psi_i^R \rangle =1$), and subject to evolutions driven by $\hat{H}^\dag$ and $\hat{H}$ respectively. The average biorthogonal position of all the eigenstates is then
\begin{equation}
    \bar{x}^{LR}(t) :=\frac{1}{N}\sum_{i}\bra{\tilde{\Psi}_i^L(t)} \hat{x} \ket{\tilde{\Psi}_i^R(t)},
    \label{eq:xlr}
\end{equation}
where the summation runs over all the states in the filled band. Note that discrete
edge states exist under the OBC, but their contribution to the summation is insignificant and thus is neglected in the subsequent discussions.
In the following, we will show that the shift of the average position over one driving cycle is quantized, and in the presence of the non-Hermitian skin effect, yields the non-Bloch Chern number defined across a generalized Brillouin zone of the momentum-time manifold.
Before discussing the pumping process, we first examine the static non-Hermitian Hamiltonian.

\subsubsection{The generalized Brillouin zone}
The generalized Brillouin zone is a generic means to label eigenstates under the OBC in non-Hermitian systems.
In a general class of non-Hermitian systems, bulk eigenstates can be significantly affected by
the boundary condition~\cite{skinrev1,skinrev2}. Under the OBC for instance, all eigenstates accumulate to the open boundaries, invalidating the conventional Bloch-wave description of bulk states. An efficient way to characterize the deformed bulk states, is to invoke the non-Bloch band theory and organize them in the generalized Brillouin zone~\cite{nhtopot4,murakami}. Indeed, such a practice is crucial for the restoration of the bulk-boundary correspondence in non-Hermitian topological models with the non-Hermitian skin effect. Specifically, following the non-Bloch band theory, we denote the static right bulk eigenstates as
\begin{align}
 \ket{\Psi_{\beta}^{R}}=&\frac{1}{\mathcal{N}}\sum_{m}\beta^m\ket{m}\otimes\ket{u_\beta^{R}},\label{eq:gbz1}
\end{align}
where $\mathcal{N}$ is the normalization factor that ensures $\langle \Psi_{\beta}^{R} | \Psi_{\beta}^{R} \rangle = 1 $, and $\ket{u_\beta^{R}}$ are the eigenstates of  $\hat{H}(\beta)$. Here the non-Bloch Hamiltonian $\hat{H}(\beta)$ in the generalized Brillioun zone is obtained by making the substitution $e^{ik} \rightarrow \beta \in \mathbb{C} $ in the Bloch Hamiltonian $\hat{H}(k)$ under the PBC. While $\beta$ can be calculated from the eigen problem of the system under the OBC, all the $\beta$ from different eigenstates typically form a closed loop on the complex plane, dubbed the generalized Brillouin zone. For non-Hermitian models without the non-Hermitian skin effect, the trajectory of $\beta$ is reduced to a unit circle on the complex plane, recovering the conventional Brillouin zone.

Based on the insight above, we adopt the parameterization $\beta=|\beta|e^{i\theta}$, where $\theta\in[0,2\pi)$ labels eigenstates in the generalized Brillouin zone, just as the quasimomentum $k$ in the conventional Brillouin zone.
The right bulk eigenstates are then labeled as $\ket{\Psi_{\theta}^{R}}$. We further introduce the left eigenstates $\ket{\Psi_{\theta}^{L}}$, which satisfy the biorthogonal relation
\begin{equation}
    \langle{\Psi^L_{\theta'}}|{\Psi^R_{\theta}}\rangle = 2\pi \delta(\theta'-\theta).
\end{equation}
Neglecting the few discrete edge states,
we can express the average biorthogonal position as an average over the generalized Brillouin zone
\begin{align}
   \bar{x}^{LR}=\frac{1}{N}\sum_{i}\bra{\Psi_i^L} \hat{x} \ket{\Psi_i^R}
  =\int \frac{d\theta}{2\pi} \bra{\Psi_\theta^L}\hat{x}\ket{\Psi_\theta^R},   \label{eq:xlrnew}
\end{align}
where the summation (over $i$) runs over all the eigenstates in the filled band.

We then define the biorthogonal Wannier states
\begin{equation}\label{OBCWannierdef}
    \left|w_{j}^{L,R}\right\rangle:=\frac{1}{\sqrt{N}} \sum_{\theta} e^{-i j \theta}\left|\Psi_{\theta}^{L,R}\right\rangle,
\end{equation}
which may not be localized in the bulk, but can instead accumulate toward the boundary due to the non-Hermitian skin effect. It is straightforward to show that the Wannier states satisfy biorthogonality, with $\langle w^L_{i}|w^R_j\rangle=\delta_{ij}$. It follows that
\begin{align}
\bar{x}^{LR} = \frac{1}{N}\sum_{j}\left\langle w_{j}^{L}|\hat{x}| w_{j}^{R}\right\rangle,\label{barx=argWC}
\end{align}
recovering the result of $\Delta \bar x^{LR}$.

Next, we consider the geometric implication of the biorthogonal Wannier center
defined through the Wannier states in Eq.~(\ref{OBCWannierdef}).
For simplicity, we consider the case where the generalized Brillouin zone is circular, leaving detailed discussions of the non-circular case in Appendix~\ref{apB}. We note that the two cases lead to the same result of quantized pumping.

In the case of a circular generalized Brillouin zone, the modulus of $\beta$ is fixed.
The left eigenstates then satisfy $\ket{\Psi^L_{\theta}} \propto \sum_m (\beta^{*})^{-m}\ket{m}\otimes \ket{u_\theta^L}$, with $\ket{u_\theta^L}$ denoting the eigenstates of $H^\dagger(\beta)$ and satisfying $\langle u_\theta^L|u_{\theta}^R\rangle =1$. We then have
\begin{align}
&\left\langle w_{j}^{L}|\hat{x}| w_{j}^{R}\right\rangle \nonumber\\  =
 &i\sum_m \int \frac{d\theta d\theta'}{(2\pi)^2}e^{im(\bar{k}-\bar{k}')}e^{ij(\theta'-\theta)} \nonumber\\
\times &\left( -ij\langle u_{\theta'}^L | u_{\theta}^R \rangle  -m\frac{\partial\Gamma}{\partial\theta}  \langle u_{\theta'}^L | u_{\theta}^R \rangle
 +\langle u_{\theta'}^L |\partial_\theta| u_{\theta}^R \rangle   \right).\label{argWC}
\end{align}
In the derivations above, we have defined $\bar{k} := -i\ln{\beta}$ and $\Gamma:=\text{Im} \bar{k}$. Because the generalized Brillouin zone is circular, we have $\bar{k}-\bar{k}'=\theta-\theta'$ and $\frac{\partial \Gamma}{\partial \theta}=0 $. It follows that
\begin{equation}
    \langle w_j^L |\hat{x} |w_j^R\rangle = j + i\int \frac{d\theta}{2\pi} \langle u_{\theta}^L |\partial_\theta| u_{\theta}^R \rangle,
\label{cirlcularWC}
\end{equation}
which is formally similar to Eq.~(\ref{HermitianWannierpump}) and Eq.~(\ref{eq:wc_LR_PBC}), where the quantity $i  \int \frac{d \theta}{2\pi } \left\langle u_{\theta}^{L}\left|\partial_{\theta}\right| u_{\theta}^{R}\right\rangle$ is identified as the non-Bloch Berry phase.

\subsubsection{Pumping under the OBC}
We are now in a position to examine the pumping process. We denote the
instantaneous left and right eigenstates as $\ket{{\Psi}_\theta^{L,R}(t)}$, with $\hat{H}(t)\ket{{\Psi}_\theta^{R}(t)}=\epsilon_\theta(t)\ket{{\Psi}_\theta^{R}(t)}$ and $\hat{H}^\dagger(t)\ket{{\Psi}_\theta^{L}(t)}=\epsilon^*_\theta(t)\ket{{\Psi}_\theta^{L}(t)}$, respectively.
Assuming adiabaticity, the time-evolved states are
\begin{align}
\ket{\tilde{\Psi}_\theta^{R}(t)}&=\exp[{i\kappa(\theta,t)}]\ket{{\Psi}_\theta^{R}(t)},\\ \ket{\tilde{\Psi}_\theta^{L}(t)}&=\exp[{i\kappa^*(\theta,t)}]\ket{{\Psi}_\theta^{L}(t)},
\end{align}
where
$ \kappa(\theta,t)  = \int_0^t d\tau [i\left\langle{\Psi_\theta^L(\tau)}|{\partial_{\tau}{\Psi}_\theta^R(\tau)}\right\rangle-\epsilon_\theta(\tau)]$.

Following the definition in Eq.~(\ref{eq:xlr}),
the average displacement over one cycle is
\begin{equation}
    \Delta \bar{x}^{LR} = \bar{x}^{LR}(T)-\bar{x}^{LR}(0).\label{eq:result_OBC}
\end{equation}

In the case that the generalized Brillouin zone remains circular during the pumping process, according to Eq.~(\ref{cirlcularWC}), we obtain
\begin{equation}
   \bar{x}^{LR}(t) = \frac{1+N}{2} + i\int \frac{d\theta}{2\pi} \langle u_{\theta}^L(t) |\partial_\theta| u_{\theta}^R(t) \rangle,\label{xlr=berryphase}
\end{equation}
where the first term is a time-independent constant, and $|u_{\theta}^{L}(t) \rangle$ and $|u_{\theta}^{R}(t) \rangle$ are respectively the instantaneous eigenstates of $\hat{H}^\dag(\beta,t)$ and $\hat{H}(\beta,t)$.
It follows that
\begin{align}
    \Delta \bar{x}^{LR} =\int_0^T dt~ \partial_t \bar{x}^{LR}(t) = C_{\text{nB}},\label{mainresult}
\end{align}
where $C_{\text{nB}}$ is the non-Bloch Chern number calculated over the manifold spanned by the generalized Brillouin zone and time
\begin{equation}
    C_{\text{nB}}=\frac{1}{2\pi i}\int_0^T dt\int d\theta ~\epsilon^{ij}\langle\partial_i {u}_\theta^{L}(t)|\partial_j {u}^R_{\theta}(t)\rangle,
\end{equation}
where $\epsilon^{\theta t}=-\epsilon^{t \theta} = 1 $. Hence, the biorthogonal displacement is still quantized.

In the case where the generalized Brillouin zone remains circular, the biorthogonal displacement of an arbitrary Wannier center in the bulk gives the non-Bloch Chern number according to Eqs.~(\ref{cirlcularWC})(\ref{xlr=berryphase})(\ref{mainresult}). In the more general case where the generalized Brillouin zone is non-circular,
the displacement of a single Wannier center, as in Eq.~(\ref{cirlcularWC}), is no longer quantized to the Chern number. Nevertheless, as we show in Appendix~\ref{apB}, Eq.~(\ref{mainresult}) is still valid in this case.

\subsubsection{Displacement without the biorthogonal formalism}
Instead of the average position under the biorthogonal construction, one might be tempted to
consider the average position using only right time-evolved states
\begin{align}
\bar{x}^{RR}(t)=\int \frac{d\theta}{2\pi} \bra{\tilde{\Psi}_\theta^R(t)}\hat{x}\ket{\tilde{\Psi}_\theta^R(t)}.
\end{align}
While such a construction corresponds more directly to physical measurements, we show in the following that average displacements calculated in this way are not quantized, except for special cases.

A first problem with $\bar{x}^{RR}(t)$ is that the generally different imaginary components of $\epsilon_\theta$ would give rise to distinct growth rates in different $\theta$ sectors.
These factors cannot be canceled, unlike in the biorthogonal case. Nevertheless, one can avoid such a problem by introducing a
renormalized average position
\begin{align}
\bar{x}^{\text{re}}(t)&:=\frac{1}{N} \sum_i \frac{\bra{\tilde{\Psi}^R_i(t)}\hat{x}\ket{\tilde{\Psi}^R_i(t)}}{\langle\tilde{\Psi}^R_i(t)|\tilde{\Psi}^R_i(t)\rangle} \label{eq:def_renorm_position}\\
&= \int \frac{d\theta}{2\pi} ~{}_{\text{re}}\langle{\tilde{\Psi}_\theta^R(t)}|\hat{x}|{\tilde{\Psi}_\theta^R(t)}\rangle_{\text{re}},
\end{align}
where $i$ runs over all time-evolved states in the given band, and the renormalized state $\ket{\tilde{\Psi}_\theta^R(t)}_{\text{re}}=\exp[i\text{Re}\kappa(\theta,t) ]\ket{{\Psi}_\theta^R(t)}$. It follows that
\begin{align}
\bar{x}^{\text{re}}(t)&= \sum_{j} \int \frac{d \theta^{\prime} d \theta}{(2\pi)^2 }~ {j} e^{i j\left(\theta^{\prime}-\theta\right)}\left\langle\Psi_{\theta^{\prime}}^{R}(t) | \Psi_{\theta}^{R}(t)\right\rangle\nonumber\\
&+\frac{i}{N}\sum_j \int \frac{d\theta}{2\pi} ~e^{-2j\Gamma}\langle u_\theta^R(t)|\partial_\theta |u_\theta^R(t)\rangle.    \label{eq:xrrt}
\end{align}
The expression above differs significantly from Eq.~(\ref{xlr=berryphase}).
The first term on the right-hand side of Eq.~(\ref{eq:xrrt}) is not a time-independent constant,
since $\left\langle\Psi_{\theta^{\prime}}^{R}(t) | \Psi_{\theta}^{R}(t)\right\rangle \neq 2\pi\delta({\theta-\theta'})$, and the second term features an additional exponential factor $e^{-2 j \Gamma(\theta,t)}$.
Note that the time dependence of $\Gamma$ comes from the time-dependent Hamiltonian, whose instantaneous eigenstates under the OBC lead to a time-dependent generalized Brillioun zone,
in a similar fashion to Eqs.~(\ref{eq:gbz1}).
It follows the shift of $\bar{x}^{\text{re}}(t)$ over one cycle would not give the non-Bloch Chern number.

However, in the absence of the non-Hermitian skin effect, $\Gamma=0$, and eigenstates belonging
to different $\theta$ sectors of the same band are orthogonal. It is then straightforward to show that
\begin{align}
\bar{x}^{\text{re}}(t)=\frac{N+1}{2} + i  \int \frac{d \theta}{2\pi } \left\langle u_{\theta}^{R}(t)\left|\partial_{\theta}\right| u_{\theta}^{R}(t)\right\rangle.
\end{align}
Since the generalized Brillouin zone is equal to the Brillouin zone in the absence of the non-Hermitian skin effect, the shift of $\bar{x}^{\text{re}}(t)$ over one cycle satisfies
\begin{align}
\Delta \bar{x}^{\text{re}} = \int_0^T dt \int \frac{dk}{2\pi} \Omega^{RR}_{kt}=C,
\end{align}
where $\Omega^{RR}_{kt}$ is defined by Eq.~(\ref{eq:defBerrycurv_PBC}).
Therefore, for systems without the non-Hermitian skin effect, the only difficulty in achieving a quantized pumping without resorting to the biorthogonal formalism is the exponential growth or decay, dictated by the imaginary components of the eigenenergies. For systems with a flat or nearly flat imaginary band, quantized pumping can still be observed (through the standard definition of expectation values using only right eigenstates), as shown in Sec~\ref{sec:modelB}.

\section{Examples}\label{sec:example}
In this section, we illustrate our results above using concrete examples.

\subsection{Non-reciprocal Rice-Mele model}\label{modelA}
\begin{figure}[tbp]
\centering
\includegraphics[scale=0.63]{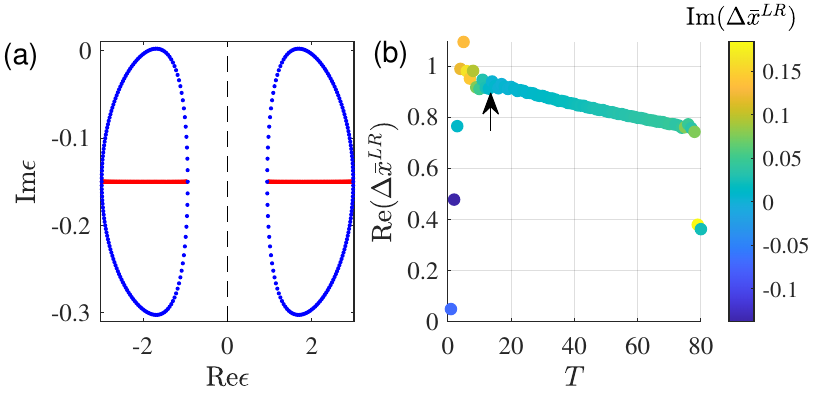}
\caption{(a) The eigenspectrum under the OBC (red) and PBC (blue) at $t=0$ for Hamiltonian (\ref{modelAH}). (b) The biorthogonal average displacement $\Delta \bar{x}^{LR}$ as a function of the pumping period $T$.
For all calculations, we take $\mu=1$ and $\gamma=0.3$. The color represents the imaginary part of $\Delta \bar{x}^{LR}$, and the arrow indicates the parameter we choose for the non-Hermitian pumping. We set $w(0)$ as the unit of energy.}
\label{adiabatic}
\end{figure}

We start with the non-reciprocal Rice-Mele model
\begin{align}\label{modelAH}
\hat{H}(t)&=\sum_{i} w(t)\left(\hat{a}_{i+1}^{\dagger} \hat{b}_{i}+\text { H.c }\right) +\sum \Delta(t)\left(\hat{a}_{i}^{\dagger} \hat{a}_{i}-\hat{b}_{i}^{\dagger} \hat{b}_{i}\right)\nonumber\\
&+\sum_{i}\left(v_{-}(t) \hat{a}_{i}^{\dagger} \hat{b}_{i}+v_{+}(t) \hat{b}_{i}^{\dagger} \hat{a}_{i}\right),
\end{align}
where $\hat{a}_i$ and $\hat{b}_i$ are respectively the annihilation operators of the two sublattice sites in the $i$th unit cell,
 $w(t)=1$, $v_{\pm}(t)=\mu+\cos (\omega t) \pm \gamma $, and $\Delta(t)=\sin (\omega t)$.
The time-dependent parameters $v_\pm(t)$ and $\Delta(t)$ undergo cyclic modulations, with the period given by $T=2\pi/\omega$. As the initial condition, we assume that the band with the smaller real eigenenergies is fully occupied at $t=0$. In Fig.~\ref{adiabatic}(a), this corresponds to the band to the left of the vertical dashed line.

First, we discuss the adiabatic condition.
In Fig.~\ref{adiabatic}(b), we show the average biorthogonal displacement $\Delta \bar{x}^{LR}$ for different driving periods $T$ of the non-reciprocal Rice-Mele model. As expected, $\Delta \bar{x}^{LR}$ only approaches the quantized value of unity in the intermediate-time regime. In the following discussions of the non-reciprocal Rice-Mele model, we always choose this intermediate time regime to ensure approximate adiabaticity.

\begin{figure}[tbp]
\centering
\includegraphics[scale=0.63]{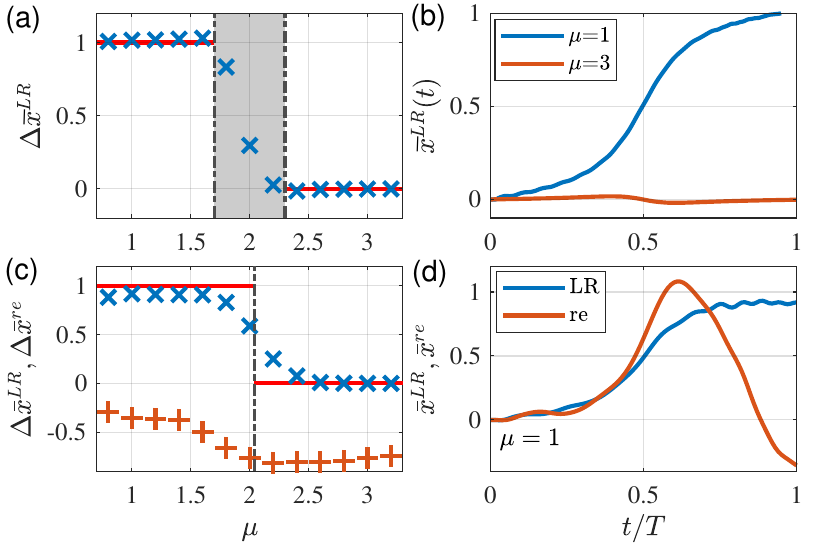}
\caption{Biorthogonal pumping of the non-Reciprocal Rice-Mele model (\ref{modelAH}). (a) The biorthogonal displacement (blue ``$\times$'') under the PBC, calculated by integrating $v_k^{LR} $ in Eq.~(\ref{eq:gv}). The red line represents the Chern number of the momentum-time manifold, and the shaded regime corresponds to a gapless region where the Chern number cannot be defined. (b) The time evolution of the biorthogonal position $\bar{x}^{LR}(t)$ at $\mu=1$ and $\mu=3$, respectively. (c) The average biorthogonal displacement $\Delta \bar{x}^{LR}$ ( blue ``$\times$"  ) and the renormalized displacement $\Delta \bar{x}^{\text{re}}$ ( red ``$+$" ) under the OBC, as functions of $\mu$. The red line represents the non-Bloch Chern number $C_{\text{nB}}$, and the phase transition occurs at $\mu=1+\sqrt{1+\gamma^2}$. (d) The time evolution of $\bar{x}^{LR}(t)$ and $\bar{x}^{\text{re}}(t)$ for $\mu=1 $ under the OBC. We take $\gamma=0.3$ for all calculations. For pumping period, we take $T=50$ in (a)(b) and $T= 15$ in (c)(d).}
\label{fig:modelA}
\end{figure}

For charge pumping under the PBC, we numerically calculate the total velocity $v^{LR}_k(t) $ using Eq.~(\ref{eq:gv}), and integrate it over the momentum-time manifold to obtain the average biorthogonal displacement {$\Delta \bar{x}^{LR}$} as shown in Fig.~\ref{fig:modelA}(a). As the parameter $\mu$ increases, the biorthogonal displacement starts quantized at $1$, but vanishes when $\mu$ becomes sufficiently large. The result is consistent with the calculated Chern number over the momentum-time manifold as shown by the red line in Fig.~\ref{fig:modelA}(a). The shaded regime corresponds to a gapless region where the Chern number cannot be defined. To provide details, in Fig.~\ref{fig:modelA}(b), we show the time evolution of the average biorthogonal position $\bar{x}^{LR}(t)=\int_0^t d\tau \int \frac{dk}{2\pi}~v^{LR}(\tau)$ for $\mu=1$ and $\mu=3$, respectively, where the quantized transport is visible.

For charge pumping under the OBC, we numerically calculate the average biorthogonal position $\bar{x}^{LR}(t)$ defined by Eq.~(\ref{eq:xlr}). Figure~\ref{fig:modelA}(c) shows the displacement of the average biorthogonal position over one cycle $\Delta \bar{x}^{LR}$ (blue). To verify the analytic results shown in Eq.~(\ref{mainresult}), we plot the non-Bloch Chern number, as shown by the red line in Fig.~\ref{fig:modelA}(c), where the phase transition occurs at $\mu=1+\sqrt{1+\gamma^2}$. For comparison, we also show the numerically calculated displacement of the renormalized position $\bar{x}^{\text{re}}(t)$ (red) defined by Eq.~(\ref{eq:def_renorm_position}).
As expected, only the biorthogonal average displacement is quantized, and consistent with the non-Bloch Chern number.
In Fig.~\ref{fig:modelA}(d), we plot the time evolution of $\bar{x}^{LR}(t)$ and $\bar{x}^{\text{re}}(t)$ to provide more details, where only the biorthogonal average displacement is visibly quantized.

\subsection{Generalized reciprocal Rice-Mele model}\label{sec:modelB}
As an example for non-Hermitian models without the non-Hermitian skin effect,
we consider a generalized reciprocal Rice-Mele model whose charge pumping was recently experimentally reported in Ref.~\cite{nc}. The time-dependent Hamiltonian can be written as
\begin{equation}
    \begin{array}{c}
\hat{H}(t)=\sum_{j}\left(J_{1}(t) \hat{b}_{j}^{\dagger} \hat{a}_{j}+J_{2}(t) \hat{a}_{j+1}^{\dagger} \hat{b}_{j}+\text { h.c. }\right) \\
+\sum_{j}\left(\left(u_{a}(t)-i \gamma_{a}(t)\right) \hat{a}_{j}^{\dagger} \hat{a}_{j}+\left(u_{b}(t)-i \gamma_{b}(t)\right) \hat{b}_{j}^{\dagger} \hat{b}_{j}\right),\label{eq:modelB}
\end{array}
\end{equation}
where $\hat{a}_i$ and $\hat{b}_i$ are respectively the annihilation operators
for the $a$ and $b$ sublattice sites of the $i$th unit cell. Following Ref.~\cite{nc}, the time-dependent parameters are
$u_{a}(t)=-u_{0} \cos (\omega t+\phi )$, $ u_{b}(t)=u_{a}(t-T / 2)$,
$J_{1}(t)=J_{0} e^{-\lambda(1-\sin (\omega t+\phi )}-\mu$,
$J_{2}(t)=J_{0} e^{-\lambda(1+\sin (\omega t+\phi )}$,
$\gamma_{a}(t)=-\gamma_{0} \Theta(x)\left(u_{a}(t)\right) \cos (\omega t+\phi )$, and $\gamma_{b}(t)=\gamma_{a}(t-T / 2)$. Here $ \Theta(x)$ is the Heaviside step function.

In Ref.~\cite{nc}, a quantized charge pumping was observed in the context of Floquet dynamics, and associated with the winding of the quasienergy around the Floquet-Brillouin zone, an intrinsically distinct topological structure in Floquet dynamics with finite period $T$.
This is curious since the experimentally observed charge pumping is generally not characterized by the biorthogonal formalism.
In the following, we will clarify this apparent discrepancy with our theory.
We note from the outset that, while the title of Ref.~\cite{nc} implies a fast (and hence non-adiabatic) pumping process that necessitates a Floquet treatment, further analysis shows that the pumping process therein is only fast in the dissipationless Hermitian limit.
In the non-Hermitian case, the adiabatic condition is modified according to Eq.~(\ref{eq:condition}) under the presence of imaginary gaps. We have checked that the non-Hermitian adiabatic condition (\ref{eq:condition}) is satisfied for all the non-Hermitian parameters and initial states in Ref.~\cite{nc}. It follows that our theory of non-Hermitian charge pumping applies.

\begin{figure}[tbp]
\centering
\includegraphics[scale=0.63]{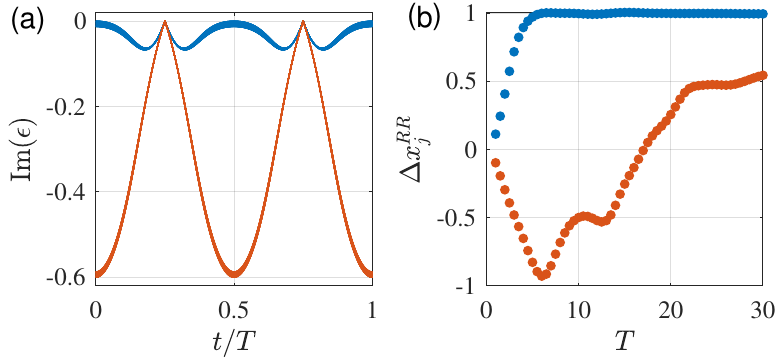}
\caption{  (a) The imaginary components of the two instantaneous bands of the time-dependent Hamiltonian (\ref{eq:modelB}). The colors are used to distinguish different bands. (b) The displacements of the conventional Wannier centers $\Delta {x}^{RR}_j$ for the two bands, for different pumping periods $T$. We take $\mu=0$, $u_0=J_0=1$, $\phi=-\pi$, $\lambda = 1.75$ and $\gamma_0=0.3$ for our calculations. We set $J_0$ as the unit of energy. }
\label{fig:Bpump_T}
\end{figure}

First, we consider the initial state and the adiabatic condition. In the experiment, the
system is initialized on a single sublattice site $a$ in a given unit cell of the bulk.
This corresponds to an equal-weight superposition of the Bloch states, and is essentially a Wannier state of the so-called ``upper'' band, the band with smaller negative imaginary components [the blue band in Fig.~\ref{fig:Bpump_T}(a)].
The model Hamiltonian is special in the sense that the imaginary components of upper-band dispersion are nearly flat, and, during the time evolution, always lie above the other band [dubbed the ``lower'' band, shown in red in  Fig.~\ref{fig:Bpump_T}(a)].
This ensures that: i) the adiabatic condition Eq.~(\ref{eq:condition})
 of the time-evolved right eigenstates in the ``upper'' band
holds, as the occupation of the ``lower'' band becomes exponentially suppressed over time; ii) the equal superposition of the Bloch states approximately holds during the time evolution. Together, these conditions dictate that the displacement calculated using only the right eigenstates is approximately quantized over a wider range of timescales.
In Fig.~\ref{fig:Bpump_T}(b), we show the displacement of the Wannier centers for different driving periods, and for systems initialized in different bands. Therein, the blue curve corresponds to the experimental observation, but when the system is initialized in lower band, the pumped charge is no longer quantized, as the adiabatic condition breaks down.

\begin{figure}[tbp]
\centering
\includegraphics[scale=0.63]{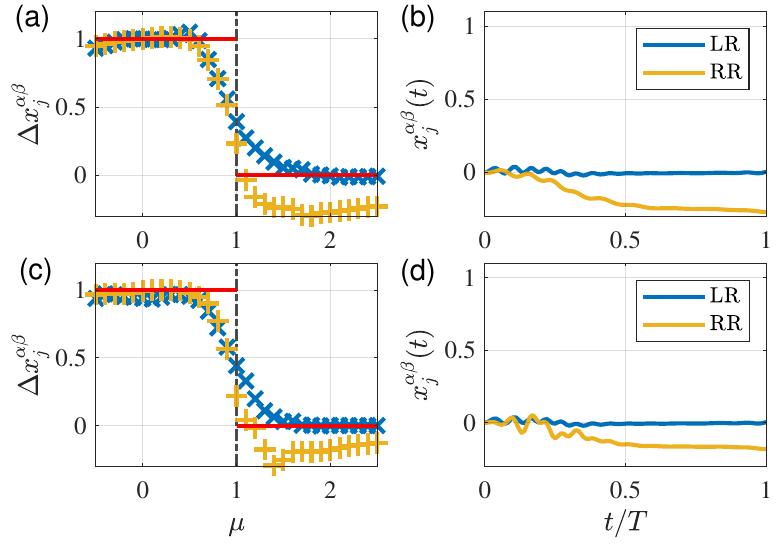}
\caption{Biorthogonal pumping for Hamiltonian (\ref{eq:modelB}). (a) The displacement of the biorthogonal Wannier center $\Delta x^{LR}_j$ (blue ``$\times$'') and that of the conventional Wannier center $\Delta x^{RR}_j$ (yellow ``$+$'') under the PBC.
The red line indicates the corresponding Chern number.
(b) The time evolution of $x^{LR}_j(t)$ (blue ) and $x^{RR}_j(t)$ (yellow) for $\mu=2$, under the PBC. (c) The displacement of the biorthogonal Wannier center $\Delta x^{LR}_j$ (blue ``$\times$'') and that of the conventional Wannier center $\Delta x^{RR}_j$ (yellow ``$+$'') under the OBC.
The red line indicates the corresponding Chern number.
(d) The time evolution of $x^{LR}_j(t)$ and $x^{RR}_j(t)$ for $\mu=2$ under the OBC . For the pumping period, we take $T=13$ in (a)(b), and $T=5|\mu|+10$ in (c)(d). Other parameters are the same as those in Fig.~\ref{fig:Bpump_T}.}
\label{modelB}
\end{figure}

For a direct comparison, we show in Fig.~\ref{modelB}(a)(b) both $\Delta x_j^{RR}$ and $\Delta x_j^{LR}$ under the PBC. While the calculated $\Delta x_j^{LR}$ agrees with the quantized Chern number (red) throughout the transition point at $\mu=1$, the quantization of $\Delta x_j^{RR}$ breaks down for $\mu>1$, where the flatness of the imaginary dispersion becomes worse.
This explicitly shows that the observation in Ref.~\cite{nc} is coincidental rather than the norm.
Finally, because the system does not have the non-Hermitian skin effect, the calculations under the OBC give similar results, as shown in Fig.~\ref{modelB}(c) and (d).

\section{Conclusion}
In conclusion, we show that quantized charge pumping in one-dimensional non-Hermitian models generally exists, but requires a biorthogonal formalism.
For this, an approximate adiabatic condition should be met, such that the time-evolved eigenstates of the system only acquire a complex phase factor.
We demonstrate that, for our biorthogonal construction, an approximate adiabatic condition exists at the intermediate time scales so that the charge-pumping problem can still be discussed.
We discuss non-Hermitian pumping for models with the non-Hermitian skin effect, and show the quantized biorthogonal pumping can be described by a non-Bloch Chern number defined in the parameter space involving the generalized Brillouin zone. Our results thus also demonstrate the utility of the non-Bloch band theory in non-Hermitian dynamics.
While the expectation values calculated with the right eigenstates correspond to physical observables in a quantum system, it is also possible to experimentally construct and detect biorthogonal expectation values~\cite{xiwang,quenchexp}. As a first attempt toward this goal, in Ref.~\cite{quenchexp}, the time-evolved left and right states are constructed from quantum-state tomography, after independently enforcing time evolutions under $H$ and $H^\dag$ on the same initial state. This enables the calculation of the biorthogonal chiral displacement, yielding the correct non-Hermitian topological invariant. While the attempt makes use of the flexible control of photonics, it is hopeful that a similar idea can be adopted to implement non-Hermitian pumping in other physical platforms.
For future studies, it is interesting to consider variants of the Thouless pumping in non-Hermitian settings, where we expect that a biorthogonal formalism should also apply.

\begin{acknowledgments}
We thank Ren Zhang for helpful discussions.
This work is supported by the National Natural Science Foundation of China (Grant No. 12374479). T. L.  acknowledges support from  the Project Funded by China Postdoctoral Science Foundation (Grant No. 2023M733719). X.-W. L. acknowledges support from the USTC start-up funding.
\end{acknowledgments}

\appendix

\renewcommand{\thesection}{\Alph{section}}
\renewcommand{\thefigure}{A\arabic{figure}}
\renewcommand{\thetable}{A\Roman{table}}
\setcounter{figure}{0}
\renewcommand{\theequation}{A\arabic{equation}}
\setcounter{equation}{0}

\section{The average biorthogonal displacement for non-circular generalized Brillouin zones}\label{apB}
In Sec.~{\ref{sec:OBC}}, we proved that the average biorthogonal displacement $\Delta \bar{x}^{LR}$ is quantized and equal to the non-Bloch Chern number when the generalized Brillouin zone remains circular. Here, we will give a more general proof, showing that the average displacement is still quantized to the non-Bloch Chern number for non-circular generalized Brillouin zones.

We start with a finite-size system in which the generalized Brillouin zone consists of discrete states labeled by $\{\beta_i~|i=1,2,\cdots,N\}$.
The corresponding right and left eigenstates are, respectively,
\begin{align}
    \ket{\Psi^R_{i}}&= \sum_{m=1}^{N}\beta_i^m \ket{m} \otimes \ket{u_{i}^R}
    :=\ket{\beta_i^R}\otimes \ket{u_{i}^R},\\
    \bra{\Psi_i^L}&=\sum_m (M)_{im}\bra{m}\otimes \bra{u_i^L}:=\bra{\beta_i^L}\otimes \bra{u_{i}^L},
\end{align}
where $\ket{u^{R,L}_i}$ are the eigenstates of $\hat{H}(\beta_i)$ and $\hat{H}^\dag(\beta_i)$, respectively. The elements of the $N\times N$ matrix $M$ are determined by the biorthogonal relation: $M_{im}=(V^{-1})_{im}/u_i^{L1}$, where $\langle u_i^L|=\left(u_i^{L1},u_i^{L2}\right)$ in the sub-lattice basis.
Without loss of generality, we consider a two-band system, and $V$ is the invertible $2N\times 2N$ matrix whose columns are the right eigenvectors of the Hamiltonian $\hat H$
\begin{align}
    V_{mi} = \beta^{m}_i u^{R1}_i,&~ V_{(m+N)i} = \beta^{m}_i u^{R2}_i,\nonumber\\
    V_{m(i+N)} = \tilde \beta^{m}_i \tilde u^{R1}_i,&~ V_{(m+N)(i+N)} =\tilde \beta^{m}_i \tilde u^{R2}_i.
\end{align}
Here $m,i =1,2,\cdots N$, and we denote the eigenstates of the two band as $|\beta_i^R \rangle \otimes | u_i^R\rangle$ and $|\tilde \beta_i^R \rangle \otimes |\tilde u_i^R\rangle$, respectively. We also denote $| u_i^R\rangle =(u_i^{R1},u_i^{R2})^{\text T}$ and $|\tilde u_i^R\rangle=(\tilde u_i^{R1},\tilde u_i^{R2})^{\text T}$ in the sub-lattice basis. Note that the matrix $V$ here is always invertible, assuming the absence of exceptional points for the non-Hermitian Hamiltonian.

In the thermodynamic limit, the index $i$ can be replaced by a continuous variable $\theta \in [0,2\pi)$. We then have $\beta=\beta(\theta)$, $\bar{k}=\bar{k}(\theta)=-i\ln\beta$, and $\Gamma=\Gamma(\theta)=\text{Im}(\bar{k})$. Since the matrix $M$ is now of infinite dimension, we make a formal substitution here:  $M_{im} \to M_{\theta m}$.

Having obtained a concrete expression for the eigenstates, we now proceed to calculate the average biorthogonal position. We first calculate $\hat{x}\ket{\Psi_\theta}$ as follows
\begin{align}
\hat{x}\ket{\Psi_\theta^R}
= &\sum_{m=1}^{N}me^{i\bar{k}m} \ket{m} \otimes \ket{u_{\theta}^R}\nonumber\\
=&-i \sum_{m=1}^{N}(\partial_{\theta}e^{i\theta m})e^{-\Gamma m } \ket{m} \otimes \ket{u_{\theta}^R}\nonumber\\
=& -i\partial_{\theta}\ket{\Psi^R_\theta}  +i \ket{\beta^R} \otimes \ket{\partial_{\theta}u_{\theta}^R}\nonumber\\
~~ &-i\sum_m m\beta^m\frac{\partial \Gamma}{\partial \theta}\ket{m}\otimes \ket{u^R_\theta}.
\end{align}
We then have
\begin{align}
    \bar{x}^{LR}=&\int \frac{d\theta}{2\pi} \langle{\Psi_\theta^L}|\hat{x}|{\Psi_\theta^R}\rangle \nonumber\\
   =& -i \int \frac{d\theta}{2\pi}\langle{\Psi_\theta^L}|\partial_{\theta}{\Psi^R_\theta}\rangle +i\int \frac{d\theta}{2\pi}\langle{u_\theta^L}|\partial_{\theta}{u_\theta^R}\rangle \nonumber\\&-i \sum_m m \int\frac{d\theta}{2\pi} M_{\theta m} \beta^m \frac{\partial \Gamma}{\partial \theta } , \label{eq:xlr=}
\end{align}
where the first term is a time-independent constant, as
\begin{align}
-i\int \frac {d\theta}{2\pi}\langle \Psi_\theta^L|\partial_\theta|\Psi_\theta^R\rangle
= \frac{1+N}{2}.
\end{align}
Here the second term in the last line of Eq.~(\ref{eq:xlr=}) gives the non-Bloch Berry phase, which cannot be obtained from a single Wannier center when the generalized Brillouin zone is non-circular. The third term looks complicated, but, as we show later, its contribution to the pumped charge vanishes in the pumping process.

Specifically, for the adiabatic evolution, both the generalized Brillouin zone and $\ket{u_\theta^{L,R} }$ become time-dependent
\begin{align}
    \bar{x}^{LR}(t) =& \frac{1+N}{2} +i\int \frac{d\theta}{2\pi}\langle{u_\theta^L}(t)|\partial_{\theta}{u_\theta^R}(t) \rangle\nonumber
    \\&-i \sum_m m \int\frac{d\theta}{2\pi} (M(t))_{\theta m} \beta^m(t)\frac{\partial \Gamma(t)}{\partial \theta } ,
\end{align}
where $\bar{x}^{LR}(t)$ is defined in Eq.(\ref{eq:xlr}), and $\ket{u_\theta^{L,R}(t)}$ are the instantaneous eigenstates of $\hat{H}^\dag(\beta,t)$ and $\hat{H}(\beta,t)$, respectively. The average biorthogonal displacement over one cycle is then
\begin{align}
 &\Delta \bar{x}^{LR} \nonumber=\int_0^T dt ~\partial_t \bar{x}^{LR}(t)
    \nonumber\\&= C_{\text{nB}} -i\sum_m m\int \frac{d\theta}{2\pi} \int_0^T dt ~\partial_t [M_{\theta m}\beta^m \frac{\partial \Gamma}{\partial \theta }].\label{eq:last}
\end{align}
Importantly, since the generalized Brillouin zone is a continuous and periodic function of the time $t$, the second term in Eq.~(\ref{eq:last}) vanishes. Hence, the average biorthogonal displacement over one pumping cycle is quantized to the non-Bloch Chern number.


\begin{thebibliography}{100}
\bibitem{Thouless1983}
D. J. Thouless, Quantization of particle transport, Phys. Rev. B \textbf{27}, 6083 (1983).

\bibitem{NiuThouless}
Q. Niu and D.J. Thouless, Quantised adiabatic charge transport in the presence of substrate disorder and many-body interaction, J. Phys. A \textbf{17}, 2453 (1984).

\bibitem{tknn}
D. J. Thouless, M. Kohomoto, M. P. Nightingale, and M. de Nijs, Quantized Hall conductance in a two-dimensional periodic potential, Phys. Rev. Lett. \textbf{49}, 405 (1982).

\bibitem{toporev1}
M. Z. Hasan and C.L. Kane, Colloquium: Topological insulators, Rev. Mod. Phys. \textbf{82}, 3045 (2010).

\bibitem{toporev2}
X.-L. Qi and S.-C. Zhang, Topological insulators and superconductors, Rev. Mod. Phys. \textbf{83}, 1057 (2011).

\bibitem{waveguide1}
Y. E. Kraus, Y. Lahini, Z. Ringel, M. Verbin, and O. Zilberberg, Topological states and adiabatic pumping in quasicrystals, Phys. Rev. Lett. \textbf{109}, 106402 (2012).

\bibitem{waveguide2}
M. Verbin, O. Zilberberg, Y. Lahini, Y. E. Kraus, and Y. Silberberg, Topological pumping over a photonic Fibonacci quasicrystal, Phys. Rev. B \textbf{91}, 064201 (2015).

\bibitem{waveguide3}
Y. Ke, X. Qin, et al, Topological phase transitions and Thouless pumping of light in photonic waveguide arrays, Laser Photonics Rev. \textbf{10}(6), 995-1001 (2016).

\bibitem{waveguide4}
A. Cerjan, M. Wang, S. Huang, K. P. Chen, and M. C. Rechtsman, Thouless pumping in disordered photonic systems, Light: Science \& Applications \textbf{9}, 1 (2020).

\bibitem{waveguide5}
O. You, S. Liang, et al, Observation of non-Abelian Thouless pump, Phys. Rev. Lett. \textbf{128}(24), 244302 (2022).

\bibitem{ultracold1}
H.-I. Lu, M. Schemmer, L. M. Aycock, D. Genkina, S. Sugawa, and I. B. Spielman, Geometrical pumping with a Bose-Einstein condensate, Phys. Rev. Lett. \textbf{116}, 200402 (2016).

\bibitem{ultracold2}
M. Lohse, C. Schweizer, O. Zilberberg, M. Aidelsburger, and I. Bloch, A Thouless quantum pump with ultracold bosonic atoms in an optical superlattice, Nat. Phy. \textbf{12}, 350 (2016).

\bibitem{ultracold3}
S. Nakajima, T. Tomita, S. Taie, T. Ichinose, H. Ozawa, L. Wang, M. Troyer, and Y. Takahashi, Topological Thouless pumping of ultracold fermions, Nat. Phys. \textbf{12}, 296 (2016).

\bibitem{fast1}
T. Kitagawa, E. Berg, M. Rudner, and E. Demler, Topological characterization of periodically driven quantum systems, Phys. Rev. B, \textbf{82}, 235114 (2010).

\bibitem{fast2}
M. S. Rudner, N. H. Lindner, E. Berg, and M. Levin, Anomalous edge states and the bulk-edge correspondence for periodically-driven two dimensional systems, arXiv preprint arXiv:1212.3324.

\bibitem{fast3}
M. C. Rechtsman, et al. Photonic Floquet topological insulators, Nature, \textbf{496}, 196 (2013).

\bibitem{fast4}
S. Malikis, and V. Cheianov, An ideal rapid-cycle Thouless pump, SciPost Physics, \textbf{12}(6), 203 (2022).


\bibitem{nc}
Z. Fedorova, H. Qiu, S. Linden, and J. Kroha, Observation of topological transport quantization by dissipation in fast Thouless pumps, Nat. Commun., \textbf{11}, 3758 (2020).

\bibitem{dissipative}
D. Dreon, A. Baumg\"{a}rtner, X. Li, S. Hertlein, T. Esslinger, and  T. Donner, Self-oscillating pump in a topological dissipative atom-cavity system, Nature, \textbf{608}(7923), 494-498 (2022).

\bibitem{molmer}J. Dalibard, Y. Castin, and K. M\o lmer, Wave-function approach to dissipative processes in quantum optics, Phys. Rev. Lett. {\bf68}(5), 580 (1992).

\bibitem{weimer} H. Weimer, A. Kshetrimayum, and R. Or{\'u}s, Simulation methods for open quantum many-body systems, Rev. Mod. Phys. {\bf93}, 015008 (2021).


\bibitem{photonics3} {\c{S}}. K. {\"O}zdemir, S. Rotter, F. Nori, and L. Yang, Parity–time symmetry and exceptional points in photonics, Nat. Mater. \textbf{18}, 783 (2019).


\bibitem{photonics4} M. A. Miri and A. Al\'u, Exceptional points in optics and photonics, Science \textbf{363}, eaar7709 (2019).



\bibitem{Non1} N. Moiseyev, Non-Hermitian quantum mechanics, {\it Cambridge University Press} (2011).
\bibitem{uedareview}
Y. Ashida, Z. Gong, and M. Ueda, Non-Hermitian physics, Adv. Phys. {\bf 69}, 3 (2020).


\bibitem{mastereqeff1}F. Song, S. Yao, and Z. Wang, Non-Hermitian skin effect and chiral damping in open quantum systems,	Phys. Rev. Lett. {\bf 123}, 170401 (2019).

\bibitem{mastereqeff2}N, Shibata and H, Katsura, Dissipative spin chain as a non-Hermitian Kitaev ladder, Phys. Rev. B {\bf 99}, 174303 (2019).
 
\bibitem{zhushiliang} P. He, Y.-G. Liu, J.-T. Wang, and S.-L. Zhu, Damping transition in an open generalized Aubry-Andr\'e-Harper model, Phys. Rev. A {\bf 105}, 023311 (2022).

\bibitem{tianyu} T. Li, Y.-S. Zhang, and W. Yi, Engineering dissipative quasicrystals, Phys. Rev. B {\bf 105}, 125111 (2022).


\bibitem{nhadia}
S. Ib\'{a}\~nez and J. G. Muga, Adiabaticity condition for non-Hermitian Hamiltonians, Phys. Rev. A \textbf{89}, 033403 (2014).

\bibitem{nhtopot4}S. Yao and Z. Wang, Edge states and topological invariants of non-Hermitian systems, Phys. Rev. Lett. {\bf 121}, 086803 (2018).

\bibitem{nhtopot5}S. Yao, F. Song, and Z. Wang, Non-hermitian chern bands, Phys. Rev. Lett. {\bf 121}, 136802 (2018).

\bibitem{murakami}K. Yokomizo and S. Murakami, Non-Bloch band theory of non-Hermitian systems, Phys. Rev. Lett. {\bf 123}, 066404 (2019).

\bibitem{nhse1}C. H. Lee and R. Thomale, Anatomy of skin modes and topology in non-Hermitian systems, Phys. Rev. B \textbf{99}, 201103(R) (2019).

\bibitem{nhse2}A. McDonald, T. Pereg-Barnea, and A. A. Clerk, Phase-dependent chiral transport and effective non-Hermitian dynamics in a bosonic Kitaev-Majorana chain, Phys. Rev. X \textbf{8}, 041031 (2018).

\bibitem{nhse3}K. Zhang, Z. Yang, and C. Fang, Correspondence between winding numbers and skin modes in non-Hermitian systems, Phys. Rev. Lett. \textbf{125}, 126402 (2020).
 
\bibitem{nhse4}N. Okuma, K. Kawabata, K. Shiozaki, and M. Sato, Topological origin of non-Hermitian skin effects, Phys. Rev. Lett. \textbf{124}, 086801 (2020).
 
\bibitem{nhse5}T.-S. Deng and W. Yi, Non-Bloch topological invariants in a non-Hermitian domain wall system, Phys. Rev. B \textbf{100}, 035102 (2019).
 
\bibitem{nhse6}Z. Yang, K. Zhang, C. Fang, and J. Hu, Non-Hermitian bulk-boundary correspondence and auxiliary generalized Brillouin zone theory, Phys. Rev. Lett. \textbf{125}, 226402 (2020).
 
\bibitem{nhsedy1}S. Longhi, Probing non-Hermitian skin effect and non-Bloch phase transitions, Phys. Rev. Research \textbf{1}(2), 023013 (2019).
 
\bibitem{nhsedy2}T. Li, J.-Z. Sun, Y.-S. Zhang, and W. Yi, Non-Bloch quench dynamics, Phys. Rev. Research \textbf{3}, 023022 (2021).
 
\bibitem{nhsedy3}S. Longhi, Unraveling the non-Hermitian skin effect in dissipative systems, Phys. Rev. B \textbf{102}, 201103(R) (2020).
 
\bibitem{skinrev1} K. Ding, C. Fang, and G. Ma, Non-Hermitian topology and exceptional-point geometries, Nat. Rev. Phys. {\bf 4}, 745 (2022).
   
\bibitem{skinrev2} R. Lin, T. Tai, L. Li, and C. H. Lee, Topological non-Hermitian skin effect, Front. Phys. \textbf{18}, 53605 (2023).


\bibitem{emilskin} F. K. Kunst, E. Edvardsson, J. C. Budich, and E. J. Bergholtz, Biorthogonal bulk-boundary correspondence in non-Hermitian systems, Phys. Rev. Lett. {\bf 121}, 026808 (2018).

\bibitem{TNag}  A. K. Ghosh, T. Nag, Non-Hermitian higher-order topological superconductors in two dimensions: Statics and dynamics, Phys. Rev. B, \textbf{106}, L140303 (2022).

\bibitem{emilapl} M. Zelenayova and E.J. Bergholtz, Non-Hermitian extended midgap states and bound states in the continuum. Applied Physics Letters, 124(4) (2024).

\bibitem{mandal} I. Mandal, Identifying gap-closings in open non-Hermitian systems by Biorthogonal Polarization, arXiv:2401.12213.

\bibitem{entropy} L. Zhou and D.-J. Zhang, Non-Hermitian Floquet topological matter—a review, Entropy, {\bf 25}, 1401 (2023).

\bibitem{xiwang}
X.-W. Luo and C. Zhang, Photonic topological insulators induced by non-Hermitian disorders in a coupled-cavity array. Applied Physics Letters, 123(8) (2023).

\bibitem{quenchexp}
K. Wang, T. Li, L. Xiao, Y. Han, W. Yi, and P. Xue, Detecting non-Bloch topological invariants in quantum dynamics, Phys. Rev. Lett. \textbf{127}, 270602 (2021).

\bibitem{classical exp 1}
M. Brandenbourger, X. Locsin, E. Lerner, et al. Non-reciprocal robotic metamaterials, Nat. Commun. \textbf{10}, 4608 (2019).

\bibitem{classical exp 2}
T. Helbig , T. Hofmann , S. Imhof , et al. Generalized bulk–boundary correspondence in non-Hermitian topolectrical circuits, Nat. Phy.  \textbf{16}(7): 747-750 (2020).

\bibitem{classical exp 3}
A. Ghatak, M. Brandenbourger, J. Van Wezel, C. Coulais, Observation of non-Hermitian topology and its bulk–edge correspondence in an active mechanical metamaterial, Proceedings of the National Academy of Sciences, \textbf{117}(47), 29561-29568 (2020).

\bibitem{RM}
M. J. Rice, and E. J. Mele, Elementary excitations of a linearly conjugated diatomic polymer, Phys. Rev. Lett. \textbf{49}, 1455 (1982).


\bibitem{nhchernPBC}
H. Shen, B. Zhen, and L. Fu, Topological band theory for non-Hermitian Hamiltonians, Phys. Rev. Lett. \textbf{120}, 146402 (2018).

\end{thebibliography}
\end{document}